\documentclass[conference]{template/IEEEtran}
\usepackage[utf8]{inputenc}
\usepackage{graphicx}
\usepackage{float}
\usepackage{color}
\usepackage{url}
\usepackage{listings}

\usepackage{color}															 
\usepackage{colortbl}	
\usepackage{varwidth}
\usepackage{amsmath}
\usepackage{multirow}
\usepackage{rotating}
\usepackage{bigdelim}

\usepackage{here} 

\usepackage[colorinlistoftodos]{todonotes}

\definecolor{azzurro}{rgb}{0.2,0.8,0.9}

\definecolor{dunkelgrau}{rgb}{0.8,0.8,0.8}
\definecolor{hellgrau}{rgb}{0.95,0.95,0.95}

\ifCLASSINFOpdf
\else
\fi
\begin{document}
%

\title{YANG2UML: Bijective Transformation and Simplification of YANG to UML}



\author{\IEEEauthorblockN{Mario Golling\IEEEauthorrefmark{1}, Robert Koch\IEEEauthorrefmark{1}, Peter Hillmann\IEEEauthorrefmark{1}, Rick Hofstede\IEEEauthorrefmark{2} and Frank Tietze\IEEEauthorrefmark{1}}
\IEEEauthorblockA{\IEEEauthorrefmark{1}Universit\"at der Bundeswehr M\"unchen,
Department of Computer Science,
85577 Neubiberg, Germany\\ Email: \{mario.golling, robert.koch, peter.hillmann, frank.tietze\}@unibw.de}
\IEEEauthorblockA{\IEEEauthorrefmark{2}Design and Analysis of Communication Systems (DACS), University of Twente, Enschede, The Netherlands\\
Email: r.j.hofstede@utwente.nl}}


%


\maketitle

\begin{abstract}

Software Defined Networking is currently revolutionizing computer networking by decoupling the network control (control plane) from the forwarding functions (data plane) enabling the network control to become directly programmable and the underlying infrastructure to be abstracted for applications and network services. 
Next to the well-known OpenFlow protocol, the XML-based NETCONF protocol is also an important means for exchanging configuration information from a management platform and is nowadays even part of OpenFlow.
In combination with NETCONF, YANG is the corresponding protocol that defines the associated data structures supporting virtually all network configuration protocols. YANG itself is a semantically rich language, which - in order to facilitate familiarization with the relevant subject - is often visualized to involve other experts or developers and to support them by their daily work (writing applications which make use of YANG).
In order to support this process, this paper presents an novel approach to optimize and simplify YANG data models to assist further discussions with the management and implementations (especially of interfaces) to reduce complexity. 
Therefore, we have defined a bidirectional mapping of YANG to UML and developed a tool that renders the created UML diagrams. This combines the benefits to use the formal language YANG with automatically maintained UML diagrams to involve other experts or developers, closing the gap between technically improved data models and their human readability.

\end{abstract}



%
\IEEEpeerreviewmaketitle

\noindent\begin{keywords}
Software Defined Networking, NETCONF, YANG, UML, Transformation, Bijective Mapping.
\end{keywords}

\section{Introduction}\label{sec:intro}
Virtualization has revolutionized IT in recent years. Due to virtualization concepts, such as cloud computing, it is possible to dynamically and flexibly adapt resources in order to face the demand of customers. Within the cloud computing paradigm, physical resources (such as CPU, memory, hard disk drives, etc.) are almost completely separated from upper layers. The consumer does not have to manage or control the underlying cloud infrastructure, but can make use of these resources~\cite{mell2011nist}. Wrt. to cloud computing, the usage can either be in the form of the so-called \textit{Infrastructure as a Service (IaaS) model} (where the consumer does not manage or control the underlying cloud infrastructure but has control over operating systems, storage, and deployed applications), the \textit{Platform as a Service (PaaS) model} (where, in addition to IaaS, the user does also neither manage nor control the underlying operating system, or storage) or the \textit{Software as a Service (SaaS) model} (where even individual application capabilities, with the possible exception of limited user-specific application configuration settings, are managed by the cloud provider)~\cite{mell2011nist}.

\subsection{Software Defined Networking vs. NETCONF}
Analogous to virtualizing computers, Software Defined Networking (SDN) allows to easier manage the network by separating the functional level, which decides where the data is sent to (\textit{control plane}), from the underlying system, which forwards the data to the selected destination (\textit{data plane})~\cite{mckeown2009software,onf2012software}. 

Besides OpenFlow, which is currently attracting significant attention from both, academia and industry~\cite{nunessurvey}, and resulted in the creation of the Open Network Foundation~\cite{openflow_switch_specification}, an industrial-driven organization, to promote SDN and to standardize the OpenFlow protocol~\cite{mckeown2008openflow}, the Network Configuration Protocol (NETCONF)~\cite{rfc6241} is also closely linked to the general topic.
NETCONF was originally designed as a management protocol for modifying the configuration of network devices~\cite{nunessurvey} by the the IETF Network Configuration Working Group~\cite{Enns06} and is a formal application programming interface (API) that allows configuration data information to be retrieved and manipulated~\cite{openflow_netconf_yang}.
Although NETCONF does not separate control and data planes (and thus does not comply with the definition of SDN as promoted by the Open Networking Foundation)~\cite{openflow_switch_specification,nunessurvey}, it shares many ideas with OpenFlow~\cite{openflow_netconf_yang}. Even more, in version 1.1, the OpenFlow Configuration Protocol (OF-CONFIG), whose task is to enable the remote configuration of OpenFlow datapaths, explicitly demands that the devices must implement NETCONF~\cite{openflow_netconf_yang}. By using a logically centralized Configuration Point and viewing the whole network as a single virtual switch, the building blocks of NETCONF and OF-CONFIG are very similar~\cite{openflow_netconf_yang,levin2012logically}. 
In addition, besides Google (as being one of the few exceptions), not many users or vendors that consider OpenFlow are willing to implement SDN networks without NETCONF~\cite{openflow_netconf}. Instead, most of the users would like to retain the field-proven smarts of their networking devices and augment them with additional functionality configured through OpenFlow~\cite{openflow_netconf}. 

In a real-life network, we will thus need both NETCONF, to configure the existing software running on networking devices, and potentially OpenFlow, to add new functionality where needed~\cite{openflow_netconf}. 
Following these considerations, this publication assist developing and maintaining NETCONF solutions, as (i) NETCONF already has a strong presence on the market (many devices already support NETCONF), (ii) NETCONF is demanded by the customers and since (iii) NETCONF is part of OpenFlow.

\subsection{NETCONF and YANG}
\begin{figure}[!t]
	\centering
	\includegraphics[width=6.5cm]{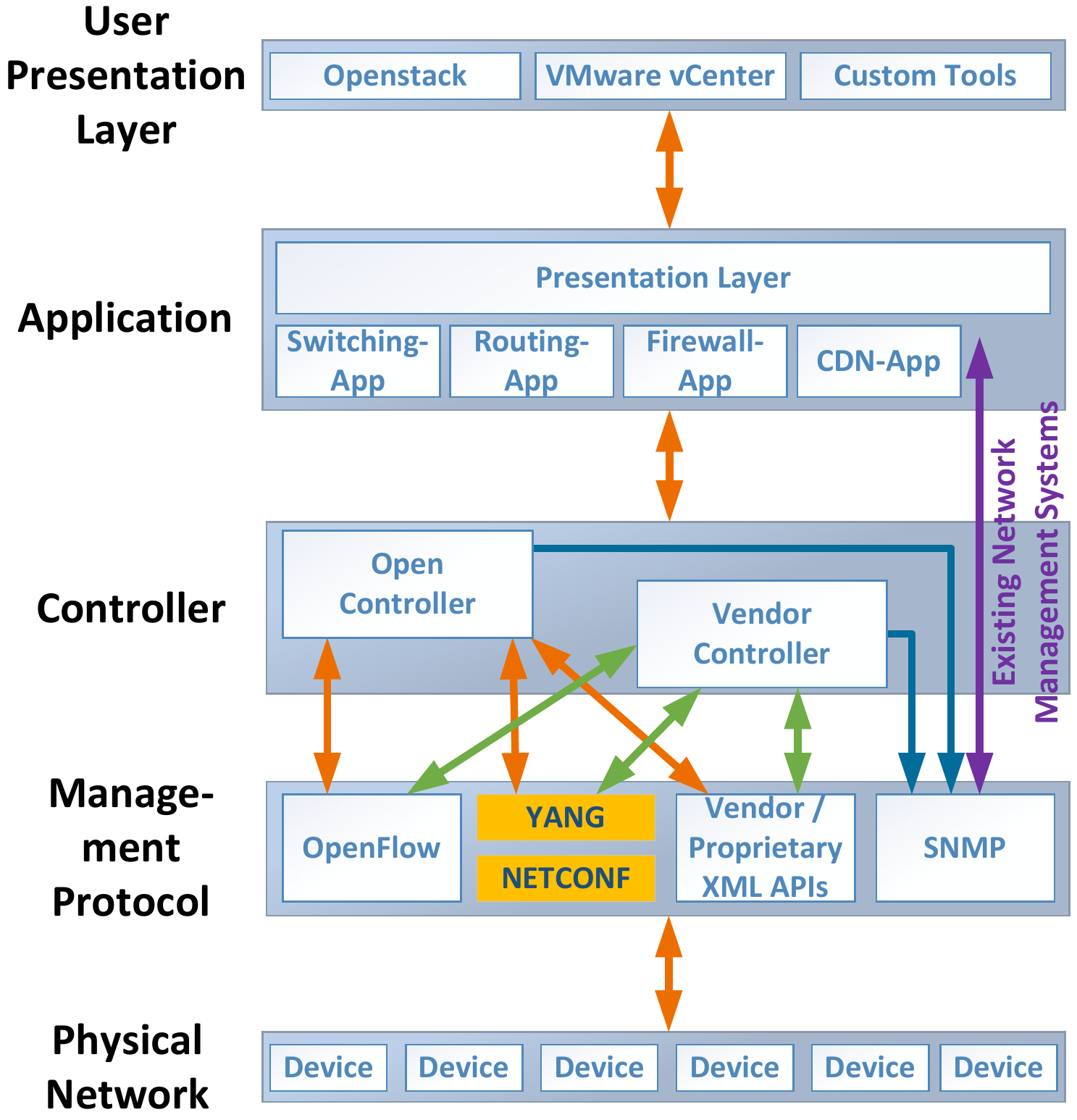}
	\caption{Classification of NETCONF and YANG in the context of SDN~\cite{sdn:netconf}.}
	\label{fig:introduction}
\end{figure}

In order to replace former solutions such as the Simple Network Management Protocol (SNMP), the Common Open Policy Service (COPS), Web-Based Enterprise Management (WBEM), the Transaction Language 1 (TL1) or CLIs, NETCONF does not include a specific standard content layer or a modeling language to be flexible~\cite{Schoenwaelder10,McCloghrie99}. Instead, the definition of data structures is done by means of YANG~\cite{Bjorklund10}, an interoperable and standardized concept for data modeling in an object-oriented fashion; YANG is short for ``Yet Another Next Generation''. 
See Fig.~\ref{fig:introduction} for an overview of the SDN approach including NETCONF and YANG. 

YANG is an extensible NETCONF data modeling language able to model configuration data, state data, operations, and notifications~\cite{juergen_netconf_aims}. As such, YANG itself is a semantically rich language supporting virtually all network configuration protocols.
Listing \ref{ietfifparts} gives an example in terms of the YANG data model for the management of network interfaces~\cite{interface}. It is expected that interface type specific data models augment this generic interfaces data model~\cite{interface}. The data model includes configuration data and state data (status information and counters for the collection of statistics).
As shown in Listing \ref{ietfifparts}, YANG uses a compact syntax, since readability has the highest priority~\cite{juergen_netconf_aims}. 
However, e.g., when writing applications (which make use of YANG), developers often would like to have YANG visualized using Unified Modeling Language (UML) diagrams for a better understanding. 
Very often, a human-readable presentation is mandatory for developers and management people for a better understanding, since these diagrams are better suited to discuss domain specific details than the textual representation of YANG, developed for abstract and automatic processing, is. In order to support this process, this paper presents a bidirectionally mappable representation model that uses the well-established YANG data model and creates an intuitive UML presentation. 
%
%
%
\begin{figure}[htb]
\lstset{caption=[Slice of the IETF YANG Module ``ietf-interfaces'']{Slice of the IETF YANG Module ``ietf-interfaces'', cf.~\cite{interface}},language=java, deletekeywords={class, if, boolean}, morekeywords={module, namespace, prefix,organization, description, container, leaf, list, type, key, revision, reference, config, mandatory}, label=ietfifparts}
\begin{lstlisting}
module ietf_interfaces_parts {
    prefix if;
    organization "IETF NETMOD Working Group";
    revision 2013-07-04 {
        description "Initial revision.";
        reference "...";}
    container interfaces {
        description "configuration parameters";
        list interface {
            key "name";
            description "configured interfaces on device"
            leaf name {
                type string;
                description "name of the interface"}
            leaf enabled {
                type boolean;
                default "true";
                description "state of the interface"
                reference "RFC 2863";}
        }
    }
    container interfaces_state {
        config false;
        description "Data nodes for the operational state";
        list interface {
            key "name";
            description "interfaces on the device"
            leaf name {
                type string;
                description "name of the interface";}
        }
    }
}
\end{lstlisting}
\end{figure}
%
%
%
Using a variety of transformation rules and by integrating the user (who is often necessary to establish the overall context), the object-oriented language ``YANG'' is transformed into the object-oriented general-purpose modeling language UML. Individual classes are reduced (e.g., two YANG classes are transformed into one UML class) to (i) reduce complexity and to (ii) improve readability. Therefore, we have defined a mapping of YANG to UML and developed a new transformation engine called YANG2UML for the automatic creation of compact object models that also renders the created UML diagrams. However, as the UML representation is intended for developers who develop their solutions on top of YANG, the corresponding UML version has to be processable by applications as well. 
Trigerstripe~\cite{Crisholm08} is such a tool to develop APIs manually by hand; this tool is used by Cisco's network management system PRIME~\cite{cisco:prime}, for example. The re-processing of the optimized UML-Models (into YANG) needs to be automated, otherwise the transformation would be too slow and will most likely not be used for further operations. As such, a reverse mapping (from the UML representation back to YANG) must always be possible in an automated fashion (i.e. without human interaction) in a unique way (\textit{bijective function}).
This combines the benefits of using the formal language YANG with automatically maintained UML diagrams to involve other experts or developers, closing the gap between technically improved data models and their readability.

\subsection{Outline of the Paper}
The remainder of this paper is structured as follows. Section \ref{sec:related-work} gives an overview of related work. 
Section \ref{sec:concept} discusses the YANG specification and our transformation policy. 
Thereafter, Section \ref{sec:poc} briefly explaines the corresponding implementation, before Section \ref{sec:evaluation} describes the evaluation and Section \ref{sec:conclusion} concludes this paper.

\section{Related Work}\label{sec:related-work}
In the area of UML-based static reverse engineering, quite a variety of solutions for re-transforming programs to UML are available~\cite{Kollman02}. 
Here, in particular, two problems can be observed. On the one hand, usually a direct conversion is performed without any optimization; thus, the model is not adapted to the specific conditions and therefore not simplified in terms of the number of objects.
On the other hand, the semantic gets lost heavily during transformation, which will infect further developments and can result in incompatibilities~\cite{Kollman02}.
For these reasons, such tools are not suitable for our use case.

In particular, the work of Schönwälder et al. has to be emphasized, too. In their work, a conceptual model to transform the Management Information Base (MIB) of the Simple Network Management Protocol (SNMP) to UML has been developed~\cite{SchoenwaelderundMueller2001}. However, because of different syntax and semantics, this work is not comparable with converting YANG to UML.

\begin{figure}[hbt]
	\centering
	\includegraphics[width=\linewidth]{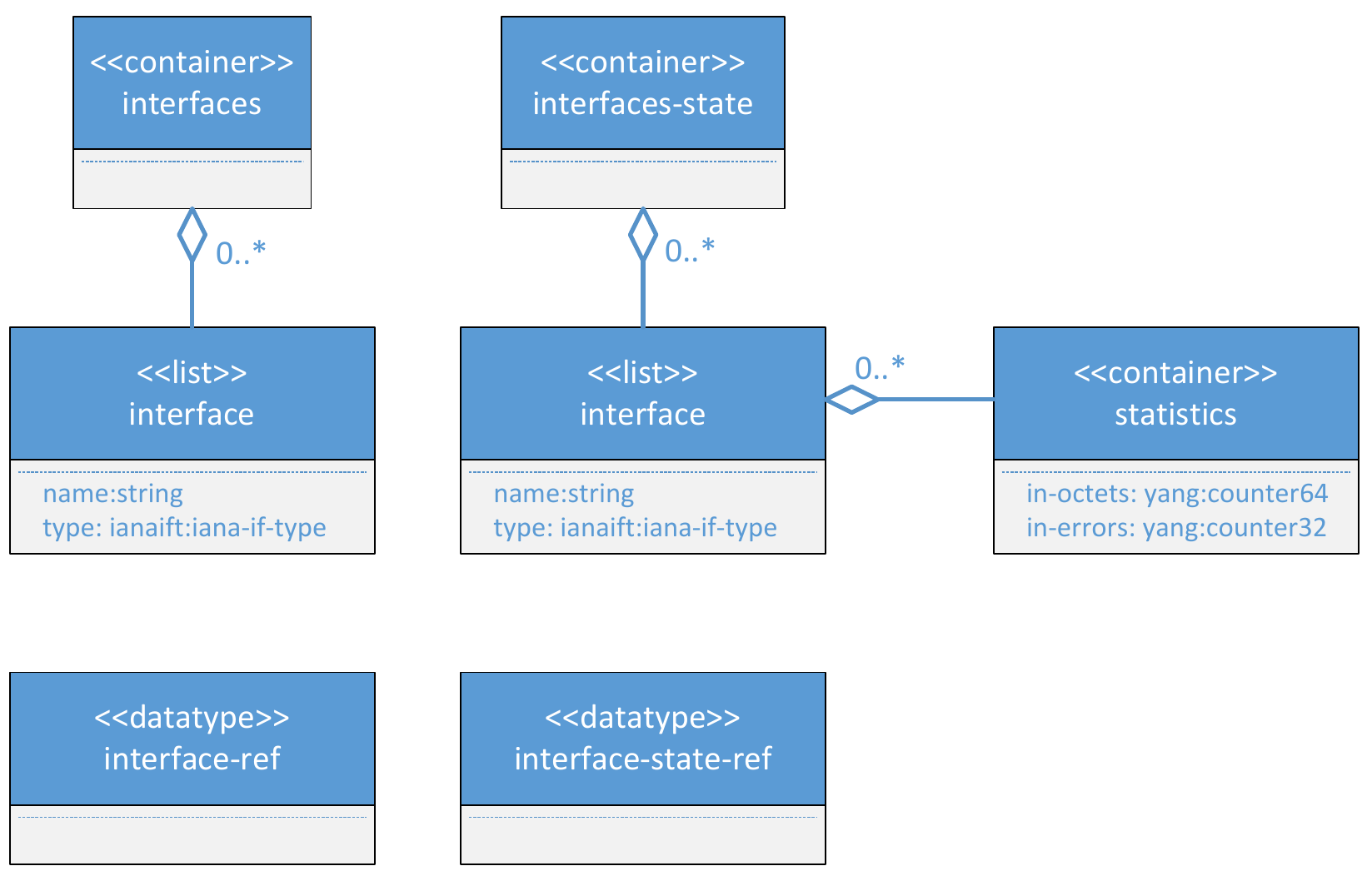}
	\caption{Result of the YANG to UML transformation when using PYANG with the Interfaces Data Model ``ietf-interfaces''}
	\label{fig:related_pyang}
\end{figure}

Finally, PYANG~\cite{pyang} has to be mentioned as the most important work in this area up to now, as it also transforms the data model of YANG directly to UML. 
Therefore, the first step of PYANG comprises a validation of YANG before within the next step YANG's compact Structure of Management Information (SMI) like syntax (cf. Listing \ref{ietfifparts}) is translated into an XML version~\cite{bray1998extensible}. The result is called YIN (YANG Independent Notation) and thus is an XML version of YANG (lossless roundtrip conversion)~\cite{juergen_netconf_aims}.
The equivalent representation of YANG information in an XML notation allows developers to use existing XML tools and tools for data filtering and validation, and thus to reduce the programming effort (e.g., see \cite{xml_tools} for an overview of XML tools).
However, since there is no reduction performed in the number of objects, the corresponding diagrams are difficult for the people to read. 
Fig. \ref{fig:related_pyang} exemplarily illustrates the result of PYANG. Here, the YANG Data Model for Interface Management (\textsc{ietf-interface}) was used. See \cite{bjorklund2013yang} for more details regarding the corresponding data model. 
Without going into more details concerning \textsc{ietf-interface}, 
it can be seen already that in this case, PYANG creates empty classes (\textsc{interfaces}, respectively. \textsc{interfaces-state}) or classes without any association (\textsc{interface-reg} or \textsc{interface-state-ref}).
%
%
%
%
%
%
%
%
%
%
%
%
As a consequence, if YANG modules grow, the corresponding UML diagrams will grow fast as well. 
%
A proposed solution to this is the use of special filters that highlight specific information in different colors to improve readability~\cite{Wallin11}. 
Since the number of objects, however, remains unchanged, this does lead to a better clarity, but is limited in its effects.

\section{Concept}\label{sec:concept}
\subsection{Top-Down vs. Bottom-Up Processing}

\begin{figure}[hbt]
	\centering
	\includegraphics[width=\linewidth]{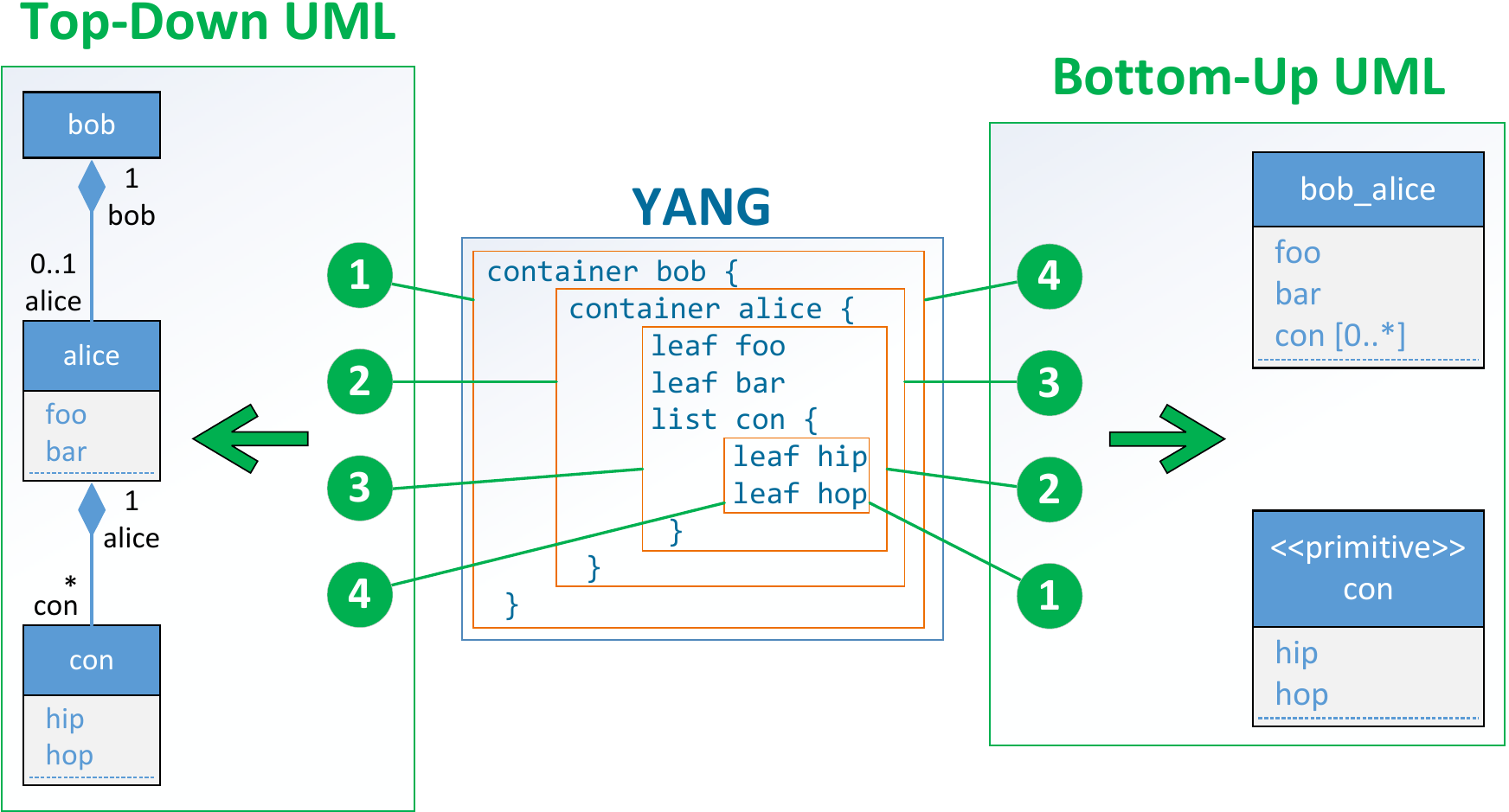}
	\caption{Top-Down vs. Buttom-Up Processing.}
	\label{fig:topdown_vs_bottomup}
\end{figure}

As the hierarchies of YANG statements are arranged as trees, there are two possibilities to process the data: Top-down respectively bottom-up.
The selected strategy and processing direction is influencing the result and therefore the quality of the data reduction process used for the optimization of the visual presentation.
Therefore, both strategies are applied and processed in parallel to combine the advantages of the respective method and to detect problems and necessities for manual decisions, 
e.g., in case of discrepancies between the two methods (see Fig.~\ref{fig:topdown_vs_bottomup}).
Processing the data with a top-down strategy enables full control of the process and an overview of the whole structure, but without having the evaluation of the (bottom) elements, which are quite important.
On the other hand, a bottom-up approach makes the elements available for further data processing, but at the expense of a more complex and costly evaluation.
To optimize the results, both strategies are combined in YANG2UML as follows:
First, the classification of elements is done by the bottom-up approach, guaranteeing an unequivocal classification of the elements.
While moving upwards in the tree, this content can be used to generate a very compact model of the YANG structure.
Second, the top-down evaluation of the tree is mandatory, because the objects created in UML are depending on each other upwardly;
e.g., no attribute can be generated without proving a class that contains the required attribute.

\subsection{Data Reduction}
For the clarity and aggregation of the visualization, a data reduction has to be performed.
As the number of objects of the model primarily depends on the number of classes and data types, trying to reduce these elements is a design requirement for the development of the set of mapping rules.
Another aspect is the consolidation and combination of data to objects, e.g., in the form of attributes.
%

The first step for the transformation, reduction and visualization of YANG datasets is a validation of the grammar and a translation to a further processable language.
RFC 6020~\cite{Bjorklund10} defines an Augmented Backus–Naur Form (ABNF) metalanguage, which will be used to validate YANG sources.
PYANG is a tool realizing this ABNF, which validates and converts YANG sources to different formats, for example YIN, UML or JSONXSL.
Therefore, we make use of PYANG to generate YIN output, which is a XML format and thus somewhat human readable but also processable with available XML libraries: \textsc{YANG} $ \xrightarrow{\text{PYANG}} $ \textsc{YIN / XML}.
Supported by that, no further validation and conversion has to be developed by us and therefore the focus can be set onto the data reduction and visualization.
%

\begin{figure}[!t]
	\centering
	\includegraphics[width=6.5cm]{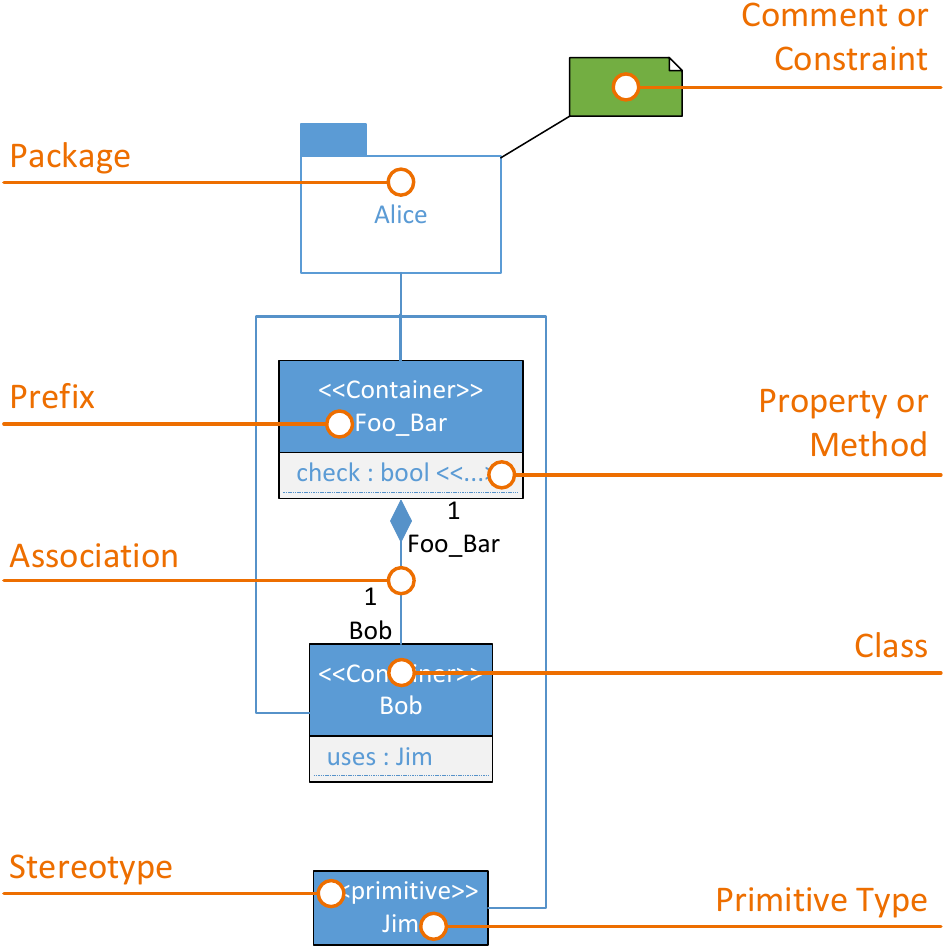}
	\caption{Transformation from YANG to UML.}
	\label{fig:yang_to_uml}
\end{figure}

A major task for the further processing is analyzing and deciding which elements can be reduced without having an influence onto the structure and semantic of the original YANG data.
Furthermore, it is essential to treat similar structures equally during the reduction process to guarantee a uniform and easy understandable picture.
This underlines the necessity of applying both strategies, bottom-up and top-down, during the analysis process: For preparing the required information, reducing the elements as well as prompting user decisions for undetermined situations.
%
A further simplification of the UML representation can be realized by using equal constructs for the mapping of YANG elements (see Fig.~\ref{fig:yang_to_uml}).
To keep the original semantic and guaranteeing a reversible transformation, stereotypes are bounded to the UML elements.
By that, the original differentiations are respected on the one hand and the presentation is easily interpretable for the user on the other hand. 
%

Another challenge is the translation of the different name definitions.
To generate a bijective mapping, the original names of the YANG structure have to be conserved.
This is realized by adding prefixes in the UML diagrams, which are generated from the names of statements and top elements.

Based on the two processing strategies, differences may appear during the conversion and the user is asked to decide the preferred conversion.
This decision has to be included within the UML model to maintain the reversibility of the mapping.
Therefore, a list of transaction rules is defined and associations are used to keep the semantic of the YANG data models and the integrity rules.
In the following, the required name conventions and transactions are described in more detail.


\begin{table*}[htb]
\begin{center}

\caption{Convention and mapping of YANG statements for visualization and data reduction.}
\label{tab:mapping}
\small
\begin{tabular}{ccc}
\hline
\rowcolor{dunkelgrau}\textsc{YANG Statement} & & \textsc{UML Element} \\
\hline
\rowcolor{hellgrau}\multicolumn{3}{c}{Name Convention} \\
Elements with same name & $ \xrightarrow[\text{6 cases}]{\text{RFC 6020}} $ & Uniqueness \\
module, submodule & $ \longleftrightarrow $ & remaining \\
extension & $ \longleftrightarrow $ & ext\_$<$name$>$ \\
feature & $ \longleftrightarrow $ & feat\_$<$name$>$ \\
identity & $ \longleftrightarrow $ & iden\_$<$name$>$ \\
typedef & $ \longleftrightarrow $ & td\_$<$name$>$ \\
grouping & $ \longleftrightarrow $ & gr\_$<$name$>$ \\
\rowcolor{hellgrau}\multicolumn{3}{c}{Removal of Illegal Characters} \\
- / \_ & $ \longleftrightarrow $ & Deletion \\
\rowcolor{hellgrau}\multicolumn{3}{c}{Mapping} \\
module & $ \longleftrightarrow $ & package, class PuK \\
submodule & $ \longleftrightarrow $ & class \\
leaf, leaf-list  & $ \longleftrightarrow $ & attribute \\
			& \rdelim\{{3}{5pt} & class \\
list, notification	& & complex datatype \\
			& 	& prefix \\
\multirow{4}{*}{container}	& \rdelim\{{4}{5pt} & no elements: improvable \\
				& & class \\
				& & complex datatype \\
				& & prefix \\
\multirow{2}{*}{grouping}	& \rdelim\{{2}{5pt} & one class: prefix \\
				& & otherwise: class \\
choice & $ \longleftrightarrow $ & state pattern \\
RPC &  $ \longleftrightarrow $ & method, complex datatype input, output \\
AnyXML &  $ \longleftrightarrow $ & attribute with comment \\
\hline
\end{tabular}
\end{center}
\vspace{-0.3cm}
\end{table*}



\subsection{Name Conventions}
Depending on the specific kind of object, every YANG element can have multiple transformations.
Because the naming of packets restricts the further use of similar names in UML, rules have to be marked by names.
The topmost namespace is built based on the \textit{module} and \textit{submodule} statements and has to be unique for the whole model, while inferior modules can have the same names.
As modules define the root of the YANG tree, they will be represented by packages in the UML visualization.
The second namespace is generated by the \textit{extension} statements, which have to be unique within a module by definition.
Therefore, if a collision appears, the name of the corresponding UML object will be expanded with the prefix \textit{ext}: \textsc{\textless{}\text{name}\textgreater{}} $\rightarrow$ \textsc{\text{ext\_}\textless{}\text{name}\textgreater{}}.
Equivalent expansions are applied for the namespaces \textit{feature}, \textit{identity}, \textit{typedef} and \textit{grouping} (\textit{feat\_}, \textit{iden\_}, \textit{td\_} and \textit{gr\_}) in cases of collisions.
All remaining statements share the last namespace, which is limited by the hierarchy of YANG:
If a collision happens, due to equal names, the upper UML prefixes are added in front of the actual name while forbidden characters are deleted.


\subsection{Mapping Rules}
The bijective mapping between basic elements of YANG and UML is realized by the rules described as follows.\\

\textbf{Module:} Modules define the root of the YANG structure; therefore they are represented by UML packages.
The respective packages can integrate subordinated elements and restrict their visibility (as in YANG, too).
Because modules on the uppermost layer in YANG can contain nodes, which are represented by attributes, complex datatypes or methods in UML, an additional class named \textit{PuK} is created to assimilate these components.\\

\textbf{Submodule:} Submodules are described by particular files in YANG, but because being dependend on superior modules, they will be integrated within the corresponding modules.
As in the case of modules, an additional class will be generated for the submodule statement, if it contains YANG elements that have to be transferred to attributes, complex datatypes or methods.
If no YANG elements exist, the statement will not be mapped to an UML object.\\

\textbf{Leaf, leaf-list:} 
The leaf statement is the smallest unit of a YANG structure; therefore it will be presented by attributes.
Also, the leaf-list statement is mapped by attributes. Because these lists can be instanciated repeatedly, the attribute obtains a cardinality $ \lbrack 0..*\rbrack$.\\

\textbf{List:} 
The mapping of the list statement can be realized by multiple UML structures, depending on the required level of reduction and the YANG elements used.
First, the analysis is done by the bottom-up strategy.
Therefore, a list (i) can be mapped to a class (complex datatypes or mixed types in the substatements), (ii) can be mapped to a complex datatype (only attributes in the substatements) or (iii) can also be transferred to a prefix in case of an object reduction (substatements contain only type definitions and/or classes).
After that, the analysis is repeated with the top-down strategy.
If solely complex datatypes and classes are found, the list will be reduced to a prefix; otherwise it will be handled as class.
The results of the two strategies are compared; if there are differences, the user will be asked for a decision.\\

\textbf{Container:} In analogy to the list statement, the container statement can be transferred in different ways.
While the semantic of the YANG elements is different, similar mapping rules can be created for list and container statements.
In case of the container statement, an additional rule has to be applied:
If a container neither has basic, nor extended or special elements, it will not contain important data.
Therefore, the container can be reduced or mapped to a class - this has to be decided by the user:
If the container may later be used for a extension of the YANG model, it cannot be removed.\\

\textbf{Grouping:}
Grouping statements are reusable, respectively referenced definitions, represented by an association in UML.
Therefore, the grouping must be an object, which can be the final point of the association.
If there is only one element within the grouping, it will be represented as a reference to reduce the number of objects.
In all remaining cases, the grouping will be represented as a class itself.\\

\textbf{Choice:}
These elements describe the selection and instantiation of exactly one of the subordinated cases.
Because of that, no UML element is able to preserve the semantic completely.
Therefore, state patterns, which are implementing a condition-based presentation, have to be used for the mapping, irrespective of the complexity of the original construct.
By that, the choice statement is realized as a superclass.
The corresponding cases are represented as classes, which are generalizations to this superclass.\\

\textbf{Notification:}
The notification statement is treated as mappings of list and container statements.
Therefore, prefixes, complex datatypes or classes are used, depending on the super- and subordinate structure.\\

\textbf{RPC:} 
In contrast to other YANG statements, RPCs describe operations with input and output parameters.
For this reason, a mapping has to be realized by generating a method in UML, which is embedded into the superior class of the module.
The input and output substatements are represented by complex datatypes.
These datatypes can be used as parameters within the respective methods.\\

\textbf{AnyXML:} 
The AnyXML statements contains XML source code which is processed during the instantiation.
Therefore, the code has to be transferred to the corresponding UML object, but a visualization as a concrete UML object is not necessary.
Consequently, the simple represenation is using an attribute, including an appended comment.

Table \ref{tab:mapping} gives an overview of name convention and ruleset for the mapping between YANG statements and UML representation.

\subsection{Descriptive Elements}
Descriptive elements specify superior elements. Therefore, no own UML objects are created. Instead, these YANG elements are assigned to the corresponding UML elements in different ways. 
Within YANG2UML, \textsc{if-features}, \textsc{must} and \textsc{when} are transferred to constraints, \textsc{type} and \textsc{default} are transferred to attributes and \textsc{config}, \textsc{key}, \textsc{mandatory}, \textsc{ordered-by}, \textsc{presence}, \textsc{require-instance}, \textsc{status} and \textsc{unique} are transferred to stereotype.
The remaining elements are mapped to comments.


\subsection{Extended Elements}
With the help of these elements, other elements can be extended by adding objects and parameters, without the need to redefine existing modules. In addition, new elements can be added easily to define new functions.
In the following, the mapping rules for these extended elements are briefly presented:\\

\textbf{Augment:} By using this statement, extensions for external modules or the module containing the augment, as well as their submodules can be made. Although many solutions are possible, within YANG2UML, augment statements are always mapped to separate classes.\\

\textbf{Extension:} 
By means of the extension statement, YANG can by extended with new constructs. The value of the statement is the new keyword, which can be used by an import in other modules.
Here, there is only one translation rule and that is using an UML class.\\
%
%

\textbf{Include and Import:} Both statements are used to extend the main module. 
Includes are used exclusively in submodules and are therefore directly integrated into the UML package of each main module. For imports, new packages are built and associated to the corresponding main module.\\
%
%

\textbf{Typedef:} 
Through typedef statements, new data types are defined for modules or submodules.
In case of an enumeration, this is represented in UML as an enumeration, too. In all other cases, the primitive data type is selected in UML.
In order to keep the data of the typedef, attributes representing the typedef are created as well.

\section{Implementation}\label{sec:poc}
\begin{figure}[!t]
	\centering
	\includegraphics[width=\linewidth]{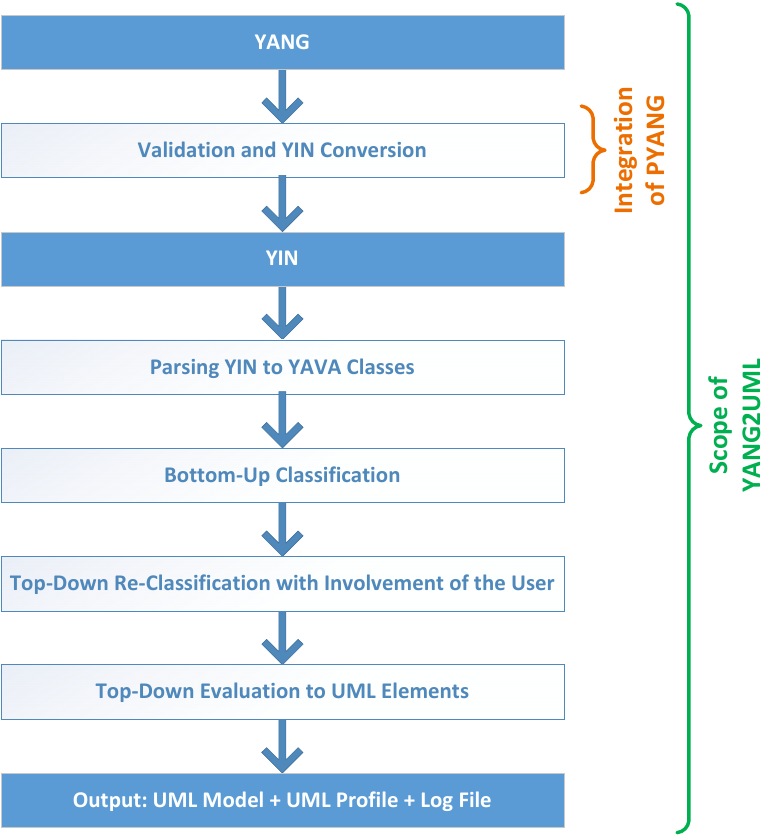}
	\caption{Steps and Artifacts of the Conversion.}
	\label{fig:steps_and_artifacts}
\end{figure}

%
%
In order to support the widest possible range of different environments, Java (with the Eclipse development environment) has been selected as the programming language for the Proof of Concept. 
Within the first step, the tool PYANG is executed with a command line call (see also Figure \ref{fig:steps_and_artifacts}). 

%
%
%
%
%

\subsection{From YIN to JAVA}
For further processing in Java, the YIN files will be parsed in again to have the possibility of accessing different elements in a convenient way. Since YIN files have a XML format, the Java-integrated XML library \textsc{javax.xml.parsers.DocumentBuilderFactory}, an API for converting XML documents into Document Object Model (DOM) object trees, will be used. So, by importing the libraries \textsc{java.wrc.com.*}, each element of the tree can be accessed very comfortable.
Within each class, the parser looks for substatements of RFC 6020~\cite{Bjorklund10}. In the first step, the names of the elements are parsed to adapt them to the corresponding UML specifications (see Table \ref{tab:mapping}).
Within the next step, the typedefs are processed as they are a prerequisite for the processing of elements that access these types. Only then all other statements can be parsed and thus the complete structure is built up. So, all objects are classified and sorted according to their statements. 
Another important aspect is the changing of names (i) that occur more than once and/or (ii) which can not be represented in UML objects. 
As YANG and UML have different namespaces, to this end, objects in UML are provided with a prefix if necessary.
As a result, all elements are imported to Java and can by transformed to UML by the reduction algorithm.

\subsection{Reduction Algorithm}
\begin{figure*}[tbp]
	\centering
	\includegraphics[width=\linewidth]{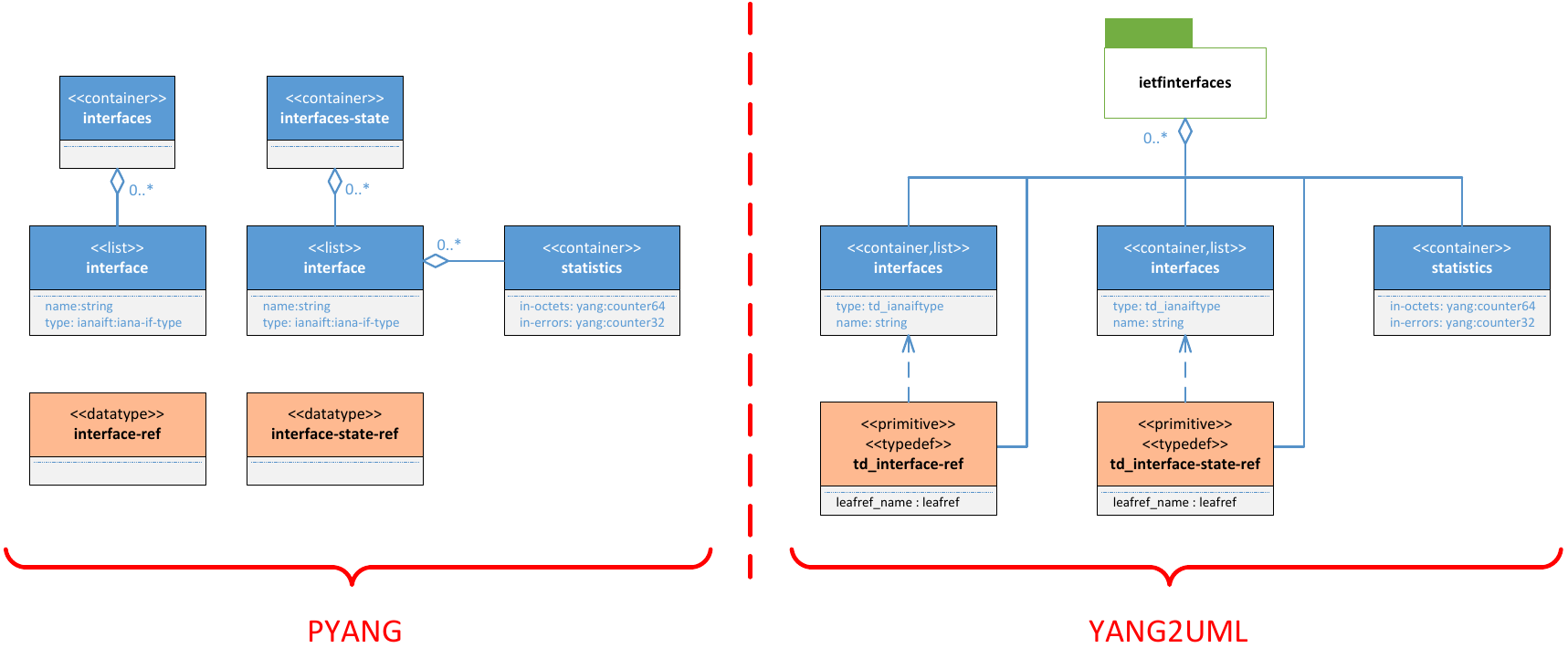}
	\caption{Direct comparison of the output of PYANG and YANG2UML for ``ietf-interfaces''}
	\label{fig:comparison_pyang_yang2uml}
\end{figure*}
As already explained, for a proper analysis, a combination of both, bottom-up and top-down, is useful. To obtain the benefits of both, each technique is used in succession. Directly after bottom-up and during the conversion to UML elements, the top-down analysis takes place. Here, it should be noted that the top-down analysis uses the information gained from the bottom-up analysis to classify the elements and involves the user in case of non-uniqueness.
In addition, various associations exist within the object model, which can only be created after the complete modeling of the objects has been performed. 
As the objects are generated gradually in one layer and are therefore not staggered in different layers, 
it is tested during the analysis which association is required.
When this is determined, the association is created and forwarded to the very top of the root of the modules or submodules.
The corresponding template is our self-developed Java class \textsc{SAssociation}, which has different constructors for the instantiation, since the startup object of the association can be a class or a primitive data type. As already mentioned, at the end, the template is forwarded to the root, so that the objects are connected with the specific association.
%
%
%
%
%

\subsection{UML Output}
For the UML output, the UML 2 API of the Eclipse development environment is used. By means of this, the individual elements of UML including the necessary associations are created. The output of the YANG2UML program consists of two different parts of UML. Firstly, the overall model is produced, which contains the main module as well as all includes and imports. Furthermore, all properties of individual objects are specified in this UML file. 
The second UML file contains the UML profile and is needed for visualization. Here, all stereotypes, built-in types and additional information are defined.

\section{Evaluation}\label{sec:evaluation}
The applicability of the generated models (for the programming of interfaces) and thus the performance of  YANG2UML can be measured by using metrics. Figure \ref{fig:comparison_pyang_yang2uml} shows the direct comparison of the output of PYANG and YANG2UML for the \textsc{ietf-interfaces} module. 
Already at first sight, it can easily be seen that now all UML objects are related to each other; therefore the depth of \textsc{ietf-interfaces} was reduced from 5 to 1, since now all objects remain in one layer. As also depicted, the number of classes (blue color) was reduced from 5 to 3, while the number of types remained constant (orange color). As a logical consequence, the number of attributes slightly increased (from 29 to 33). 
Table \ref{tab_quantitative_eval} gives an overview of several examples (including \textsc{ietf-interfaces}). All of these show that the complexity of the generated models is significantly lower, compared to PYANG. Especially the statistics from the module \textsc{ietf-routing} show impressively that the number of objects could be reduced by more than half. 

\begin{table*}[ht]
\begin{center}
\caption{Quantitative evaluation of PYANG and YANG2UML}
\label{tab_quantitative_eval}
\small
\begin{tabular}{lcccccc}
\hline
\cellcolor{dunkelgrau}\textsc{Module} & \cellcolor{dunkelgrau}\textsc{Depth} & \cellcolor{dunkelgrau}\textsc{Classes} & \cellcolor{dunkelgrau}\textsc{Types} & \cellcolor{dunkelgrau}\textsc{Attributes} & \cellcolor{dunkelgrau}\textsc{References}\\

\hline
\textsc{ietf-interfaces} - PYANG 					& 5 & 5 & 2 & 29 & 21\\
\textsc{ietf-interfaces} - YANG2UML 				& 1 & 3 & 2 & 33 & 25\\
$ $\\
\textsc{ietf-netconf-monitoring} - PYANG 		& 5 & 30 & 1 & 29 & 21\\
\textsc{ietf-netconf-monitoring} - YANG2UML 	& 1 & 22 & 3 & 51 & 36\\
$ $\\
\textsc{ietf-routing} - PYANG 						& 5 & 63 & 0 & 64 & 48\\
\textsc{ietf-routing} - YANG2UML 				& 1 & 25 & 16 & 90 & 74\\

\hline
\end{tabular}
\end{center}
\end{table*}

Furthermore, a qualitative evaluation (by human operators) has been carried out. Here, the possibility of the user to influence the final output was also evaluated as very positive. This ensures that it can be decided individually for each module, which is the most suitable mapping and thus the usability of the program is increased. However, due to this influence on the output of the program, this in turn, of course, affects quantitative aspects.

\section{Conclusion}\label{sec:conclusion}

In today's time, the network must increasingly adapt to the ever changing demands of the customers in a dynamic way. This requires that network components can be dynamically programmed and configured. Within the development of programmable systems, however, the programming is often made by means of object models. Although a mapping of hierarchical data models (YANG) to object-oriented models (UML) is, in principle, easily possible, practice has nevertheless shown that a meaningful and semantically correct mapping poses specific challenges. Especially the fact that a language with a lot of different constructs and different semantics (YANG) has to be mapped to another language with very few constructs (UML), without completely losing/changing semantics, represents a major  problem. Although, with PYANG, a converter is already available, the results are very limited, because PYANG's UML models comprise a very large number of objects and thus a large complexity. Therefore, YANG2UML has been developed, which can reduce the number of objects, thereby reducing the degree of complexity. As the evaluation has pointed out, the conversion by means of YANG2UML is successful and the object models generated are more compact, and contain more information (as e.g., imports and includes are included as well). In addition, the program also offers scope for some extensions and customizations.

\section*{Acknowledgment}

Special thanks go to Claudia Armbruster and Martin Wandtke, who made a significant contribution for this publication with their high-grade bachelor’s thesis. In addition, the authors want to thank Cisco Systems and especially Alexander Clemm for helpful discussions, valuable comments and their entire support for this research. Without Alexander Clemm, this publication would not exist.

Furthermore, the authors wish to thank the members of the Chair for Communication Systems and Internet Services at the Universit\"at der Bundeswehr M\"unchen, headed by Prof. Dr. Gabi Dreo Rodosek as well as the Design and Analysis of Communication Systems (DACS) group of the University of Twente, Enschede, for helpful discussions and valuable comments on previous versions of this paper.
%
%

\bibliographystyle{IEEEtran}
\bibliography{ms}

\end{document}